\documentclass[preprint2]{emulateapj}
\usepackage{graphicx}
\usepackage{txfonts}
\usepackage{url}
\begin{document}

\title{Accretion in Protoplanetary Disks by Collisional Fusion}
\author{J. S. Wettlaufer}

%\altaffiltext{1}{
%\affiliation{
\affil{Department of Geology \& Geophysics, Department of Physics  and \\
Program in Applied Mathematics\\
Yale University, New Haven, Connecticut, 06520-8109, USA\\}
\affil{Nordic Institute for Theoretical Physics, 106~91, Stockholm, Sweden
\\ $ $Revision: 1.48 $ $ (\today)
}

\begin{abstract}
The formation of a solar system such as ours is believed to have followed a multi-stage process around a protostar and its associated accretion disk.  Whipple first noted that 
planetesimal growth by particle agglomeration is strongly influenced by gas drag, and Cuzzi and colleagues have shown that when midplane particle mass densities approach or exceed those of the gas, solid-solid interactions dominate the drag effect.  The size dependence of the drag creates a ``bottleneck'' at  the meter scale with such bodies rapidly spiraling into the central star, whereas much smaller or larger particles do not.    Independent of whether the origin of the drag is angular momentum exchange with gas or solids in the disk, successful planetary accretion requires rapid planetesimal growth to km scales.  A commonly accepted picture is that for collisional velocities $V_c$ above a certain threshold value, ${V_{th}} \sim$ 0.1-10 cm s$^{-1}$, particle agglomeration is not possible; elastic rebound overcomes attractive surface and intermolecular forces.
 However, if perfect sticking is {\em assumed}  for all ranges of interparticle collision speeds the bottleneck can be overcome by rapid planetesimal growth.
While previous work has dealt with the influences of collisional pressures and the possibility of particle fracture or penetration, the basic role of the phase behavior of matter--phase diagrams, amorphs and polymorphs--has been neglected.  
Here it is demonstrated for compact bodies that novel aspects of surface phase transitions
provide a physical basis for efficient sticking through collisional melting/amphorphization/polymorphization and subsequent fusion/annealing to extend the collisional velocity range of primary accretion to $\Delta V_c \sim$ 1-100 m s$^{-1} \gg {V_{th}}$, which encompasses both typical turbulent RMS speeds and the velocity differences between boulder sized and small grains $\sim$ 1-50 m s$^{-1}$.  Therefore, as inspiraling meter sized bodies collide with smaller particles in this high velocity collisional fusion regime they grow sufficiently rapidly to $\sim$ 0.1 - 1 km scale and settle into stable Keplerian orbits in $\sim$ 10$^5$ years before photoevaporative wind clears the disk of source material. The basic theory applies to low and high melting temperature materials and thus to the inner and outer regions of a nebula. 
\end{abstract}


\maketitle

\section{Introduction}\label{sec:intro}

\subsection{Cosmogonical Context}

The origin of the solar system and the formation of solar type stars are wed through contemporary studies of primitive solar nebula from a wide range of perspectives including observational and theoretical astrophysics, nucleosynthesis, planetary dynamics, accretion physics and meteoritics \citep{Whipple:64, Adachi:76, Weidenschilling:77, LissauerAR:93, Cuzzi:06, Quitte:2007, Scott:07, Dullemond:07, Blum:08, Armitage:09}.  The general scenario \citep{Armitage:09} includes a sequence of stages: (a) the initial collapse of interstellar gas nucleating the central protostar ($\sim$ 0.1 Myr) (b) slow mass accretion onto the star and primary planetesimal formation around the evolving accretion disk ($\sim$ Myr) followed by (c) a phase ($\sim$ Myr) during which the accretion rate drops significantly allowing the photoevaporative wind to ``sever'' the disk into two portions at a radius that depends on the ratio of the stellar accretion rate to the mass loss rate due to photoevaporation, and finally (d)  a rapid {\em clearing phase} ($\sim$ 0.1 Myr) during which the inner disk is accreted onto the central star and the lightest elements of outer disk are removed by direct exposure to photoevaporative UV flux.   
Radiometric dating reveals that Ca-Al-rich inclusions (CAIs) within carbonaceous chondrite meteorites are about 4.57 Gy old, thereby constituting the first planetary materials \citep[see e.g.,][]{Scott:07}. Several My later, well after protoplanetesimals had already been processed, chondrules within chondrites were still forming \citep{Scott:07}.  

Whereas very small particles move ostensibly as does the slightly sub-Keplerian nebular gas,  planet sized objects couple very weakly to the  gas.  Therefore, the build up of protoplanets reaches an impasse at the meter scale due to the exchange of angular momentum with the gas; this relative motion is responsible for a drag that causes them to drift inward and they are lost to the central star on time scales much less than the CAI to chondrule interval \citep{Blum:08, Weidenschilling:08, Armitage:09}. The competition between gravitational settling and turbulence can lead to a disk midplane enhanced by solid matter and thus the same basic impasse is operative due to solid-solid angular momentum exchange which dominates the net drag effect  \cite[e.g.,][and refs. therein]{Cuzzi:1993}.  Therefore, an outstanding problem in cosmogony is to understand how, when objects can agglomerate to the critical $\sim$ meter scale  by known {\em low} collisional velocity accretion processes \citep{Dominik:07, Blum:08, Guttler:10}, they are  not then rapidly lost into the central star?  Hence, it is sought to reveal the basic mechanisms through which matter can accrete sufficiently rapidly to slow their radial motion and thwart their demise.   

\subsection{Theories of Solid Aggregation in Disks}

\subsubsection{Gravitational Collapse}

Two general approaches are advanced; one focuses on building planetesimals from grain-grain accretion and the other proposes gravitational collapse of the disk---treated as either a one or two-component (solid and gas) ``fluid''---on length scales much larger than grain scale processes.  The latter, Safronov-Goldreich-Ward \citep{Goldreich:04} and related \citep{Youdin:02} theories, overcome the bottleneck without the need to treat the microphysics of particle agglomeration, by studying the gravitational collapse of the one (solid or gas) or two-component (solid and gas) disk.  The idea is that when the solids dominate, as matter settles to the midplane the density exceeds a critical value and self gravity induces an instability with a characteristic length scale that  produces planetesimals either sufficiently large to settle into stable Keplerian orbits or of sufficient abundance to drive further gravitational accretion eventually leading to the same large planetesimal state.  When gases dominate, collapse is predicted to form giant gaseous protoplanetesimal cores that drive further accretion of gas \citep{Boss:97}.  Gravitational collapse by the settling of solids to the midplane is criticized \citep{Weidenschilling:95, Cuzzi:06, Dominik:07} on the basis that it ignores the effects of gas pressure that drive turbulent shear and maintain the disk density below the critical value.  However, the nature of disk turbulence is invoked both to assist and suppress various types of aggregation.  For example \cite{Johansen:07} argue that the turbulence driven ``streaming instability'', associated with the relative motion between solids and gas, can localize solids in the midplane and facilitate agglomeration whereas \cite{Wilkinson:08} argue that in ``Stokes trapping'' turbulence suppresses the aggregation of solids.  

\subsubsection{Sequential Aggregation of Particles}

Agglomeration through surface effects, studied in many experiments and simulations \citep{Chokshi:93, Dominik:07, Blum:08, Guttler:10}, posits that above a certain threshold collisional velocity, ${V_{th}} \sim$ 0.1-10 cm s$^{-1}$, 
particle agglomeration is not possible; elastic rebound overcomes attractive intermolecular forces.  Additionally, the experimental agglomerates are generally of high porosity and their fragility is often cited as a central difficulty with such an approach to planetary accretion \citep{Youdin:02, Wilkinson:08}.   While, for example, van der Waals interactions can accrete highly porous bodies \citep{Dominik:07, Blum:08}, both the long nebular time scales and temperature variations will facilitate densification, and thus increased strength, through the {\em sintering} and {\em annealing} of polycrystalline matter---well known processes in condensed matter physics that has power law scaling in time \citep{Dash:06}.  Annealing and sintering are apparently not commonly discussed in regards to the issue of porosity in laboratory experiments of accretion.  Therefore, a basic understanding of surface and interfacial physics and the processes of cohesion and inelasticity is required to assess accretion scenarios and the range of validity of perfect sticking \citep{Cuzzi:06, Cuzzi:07}.  

\eject

\subsection{\label{sec:Nebular}Nebular Thermodynamic Regimes}

Outside the framework of the Minimum Mass Solar Nebula (MMN), it is a challenging exercise to calculate the relevant phase fronts in an evolving accretion disk \citep{Stevenson:88, Lecar:06, CieslaCuzzi:06, Garaud:07, Armitage:09}.  For example, depending on, {\em inter alia},  turbulent mixing, the gas density, the nebular optical depth and the epoch of evolution, the snowline can vary many AU with a temperature $\sim$ 170 $\pm$ 20 $K$, moving inward as the disk evolves.   Thus, separate from the feasibility of computing it, knowledge of the entire thermodynamic history of a particle undergoing a two-body collision is highly model dependent.  However, it is important to put the situation discussed here in thermodynamic context for particle collisions.  For a given material the basic kinetics of growth at a fixed temperature in a nebula can strongly influence its bulk and surface properties and hence collisional physics.  Based solely on equilibrium considerations, ice alone exercises an enormous area of its phase diagram in a disk.  Regardless of the material, it is well known that a host of basic equilibrium and disequilibrium phenomena will influence the bulk and interfacial properties \citep{Dash:06}.  Therefore,  depending on the detailed  history of a particle it may, over the long times associated with disk evolution, have annealed to obtain normal compact laboratory properties \citep{Dash:06}, or it may have a highly porous structure \citep{Blum:08, Guttler:10}.  These issues are discussed in section \ref{sec:Liquidity}, but we note that, in the absence of fracture, the mechanism described here provides a wide range of high sticking probability collisions for a very broad range of particle densities and thermodynamic conditions.  Hence, this new mechanism acts in concert with other mechanisms of particle destruction and sticking \citep{Guttler:10}.

\subsection{\label{sec:Overview}Overview of Results}

In this paper a new thermodynamic mechanism of mass agglomeration, termed {\em collisional fusion} is described.  The mechanism involves the confluence of collisional energetics, intermolecular interactions and structural surface and bulk phase transitions.  
While the stresses in solid/solid collisions have been embodied in a range of planetesimal studies through examination of, among other things, the restitution coefficient \citep{Chokshi:93, Bridges:96, Higa:98, Guttler:10}, their influence on the bulk and {\em interfacial phase behavior} has heretofore received little attention.  Due to a combination of accessibility and the dual role of relevance to both the formation of the gas giants and the dynamics of Saturn's rings, there have been a number of experiments on the collisions of ice particles under various conditions  \citep[e.g.,][]{Bridges:84, Bridges:96, Higa:98, Dominik:07, Blum:08}, but none at the very high pressures and low temperatures where this new mechanism is operative.   Depending on the temperature and pressure, many materials can take on multiple equilibrium crystal structures ({\em polymorphs}) and moreover, under other conditions the crystalline order of one of the polymorphs can be lost and the system becomes {\em amorphous} such as is the case with glass.  Water substance exhibits a rich polymorphism and amorphism in its phase behavior.  
Here, classical collision physics is modified by an extension of the idea of {\em damage assisted interfacial melting} \citep{Dash:06} to the very high pressures that drive the fusion of matter by momentary liquifaction/amorph-polymorphization and freezing/annealing.   
The theory is quantitatively borne out in solid ice (Ih and Ic) down to $T \sim$ 150 $K$, low density amorphous ice (LDA) and high density amorphous ice (HDA) at lower temperatures.   Moreover, the process occurs in silicon, which melts at $\sim$ 1690 $K$, and where the range of collisional fusion speeds is $\Delta V_c \sim$ 100-1000 m s$^{-1}$, as seen in the recent molecular dynamics simulations of  \cite{Suri:08}.  Finally, bodies of inhomogenous composition can fuse in this manner.  For example, a body with an icy mantle of sufficient thickness can fuse with another or other multicomponent materials, such as olivine, with similar melting temperatures can act in the same manner so long as one has access to the relevant phase diagrams.  Hence the approach provides a framework for collisions from the inner to the outer nebula.  For clarity and brevity the focus here is on the latter, where it is found that perfect sticking is extended to a range of collisional speeds $\Delta V_c \sim$ 1-100 m s$^{-1}$ which encompass both typical turbulent RMS speeds and the velocity differences between boulder sized and small grains $\sim$ 1-50 m s$^{-1}$ \citep{Johansen:07}.  

The development of the paper is as follows.  In section \ref{sec:Summ} the basic physical processes at play and an outline of the logic structure of the mechanism of collisional fusion is summarized. The detailed treatment of the thermodynamic state theory of collisional fusion is described in  \ref{sec:Thermo} after which, in \ref{sec:Drift} the ideas are used to address the bottleneck problem by calculating the fate of drifters beginning from several places in nebulae experiencing a wide range of turbulent enhancement/suppression of midplane solids.  A guide to the variables and symbols used is provided in table \ref{table:Variables}.  Conclusions are drawn in \ref{sec:Conclude}. 

\section{Summary: Collisional Fusion and Accretion}
\label{sec:Summ}

\subsection{Heuristic Overview: Adhesion, Bouncing, Fusion}

Aggregation of snow on the ground or dustballs on a floor is facilitated by weak interparticle collisional speeds and attractive intermolecular forces such as van der Waals interactions.   If the temperature does not rise above freezing snow crystals will fuse together ({\em sinter}) and then densify through annealing/coarsening driven by the tendency to reduce surface energy.  Such processes are driven by either surface or volume diffusion or the transport of mobile surface films \citep{Dash:06} and they underlie the fact snow drifts soon form a crust on their surfaces or that one can form a solid metal object beginning with a powder \citep{Herring:51}. Hence,  the agglomeration and densification of matter is facilitated by time and very low collisional speeds where weak attractive interactions dominate.  However, when we toss an ice cube against the wall it will bounce, falling to the floor and often shattering.  Throwing it at a higher velocity results in shattering upon collision with the wall.  The theory developed quantitatively here considers the situation in which the wall is made of ice, and an ice ball is thrown at it some 10's of ms$^{-1}$. Upon contact, the interfacial pressure rises dramatically and some of the collisional energy momentary liquifies or disorders a thin region shared by the two surfaces.  In a nebula the ambient temperatures are sufficiently low relative to the melting points of materials under consideration that any liquid (or disordered material) will rapidly freeze (or anneal) thereby fusing the particles so long as their stored elasticity is not so rebound occurs first.  Thus, there is a low speed fusion and annealing of molecules and small particles, a higher speed bouncing and then an even higher speed--reentrant--fusion.  This basic process is not specific to ice, but ice is particularly relevant to nebulae and holds common terrestrial experience.  The process of high speed collisional fusion simply depends on the phase diagram of the material and the place (and hence thermodynamic conditions) in a nebula where a collision occurs.  It can operate, {\em mutatis mutandis}, in the inner or outer nebular regions and in all manner of materials.  The bottleneck problem is addressed by analyzing a meter sized midplane object rapidly spiraling into the central star and colliding with a range of smaller particles down to a fraction of a centimeter.   The growth rate of the inspiraling object is calculated using the theory of collisional fusion and a disk model that includes a parameterization of turbulence to estimate the midplane particle density.  It is found that the inspiraling object grows sufficiently rapidly by this mechanism to settle into a stable Keplerian orbit. 

\subsection{Collisional Fusion: The Sequence of Processes}

This basic mechanism is not specific to ice, but to make concrete predictions the theory is demonstrated with calculations that focus on ice in two nebular regions beyond the snowline.  The logic sequence is as follows.  (i) Upon collision the interfacial pressure rises dramatically.  The associated shift in the melting transition, or structural phase transition to an amorph/polymorph is determined from the phase diagram of \cite{Straessle:07} shown in Figure \ref{fig:PhaseDiagram}.  (ii) The fraction of the collisional energy/area inducing damage and hence the associated interfacial liquidity/disorder is determined from Equation \ref{eq:film} and shown in Figure \ref{fig:d(V)}.  (iii)  
The  laboratory experiments of  \cite{Higa:98} are used to constrain the degree of interfacial damage, $\xi$, as shown in Figure \ref{fig:V(crit)} below.  These are extended to the astrophysical regime by harnessing the low temperature high pressure phase behavior determined from the laboratory experiments  of \cite{Mishima:96} and \cite{Straessle:07}.   
(iv) If the time to refreeze (anneal) the liquid (amorph/polymorph) so produced is of order or less than the collision time then the particles fuse together.  There is a critical velocity beyond which the collision time is too rapid and fusion fails.  Thus, agglomeration is reentrant with collision speed $V_c$.  Fusion occurs when ${V_{U}} \ge V_c \ge {V_{L}} \gg$ ${V_{th}}$ and for ice a new high speed window of perfect sticking is found; $\Delta V_c = {V_{U}}-{V_{L}} \sim$ 1-100 m s$^{-1}$.  This range is extended to larger $V_c$ for higher melting temperature materials. 
Therefore,  the degree of damage induced disorder, and the temperature and particle size dependent collision versus freezing times, combine to determine the new collisional fusion window $\Delta V_c$ shown in Figure \ref{fig:V(range)}.

\section{Interfacial Thermodynamics and Phase Behavior}
\label{sec:Thermo}

\subsection{Thermodynamic Interfacial Damage State Theory}

The ubiquity of gravitational and intermolecular forces is responsible for their different, but
omnipresent, influence on all macroscopic bodies. They are both power law forces and they
both influence the structure and properties of materials. However, due to their vastly different
strengths and ranges of interaction they influence matter on
disparate length scales. While we have a more visceral intuition for the fact that the gravitational field layers the structure of planetary and stellar atmospheres and interiors, there is a much wider range of phenomena wherein intermolecular forces between media influence their behavior \citep[e.g.,][]{Amit, Dietrich}.  For example, liquids less polarizable than the substrates upon which they stand will ``wet'' the surface by spreading into a thin film.   

When studying the phases of matter one appeals to bulk phase diagrams.  However,  the presence of external fields (e.g., electric, magnetic, gravitational, intermolecular) can shift the region of phase stability substantially such as exists in the interior of a star due to gravitationally enhanced pressure. So it is that the presence of intermolecular force fields across a surface shift the equilibrium states of matter from bulk coexistence and can stabilize the liquid phase in the solid region of the bulk phase diagram \citep{Dash:06}. This so-called {\em interfacial premelting} occurs in all classes of materials; metals, rare gases, semiconductors, quantum solids and molecular solids, including ice wherein there are a host of astrophysical and geophysical consequences \citep{Dash:06}.  For the present development, while I will often refer to liquid, it should be viewed as a {\em synonym} for {\em the available high density phase} which can be amorphous or another polymorph of the solid as discussed in section \ref{sec:Liquidity}.  

The concept of damage assisted interfacial melting was developed by  \cite{Dash:01} to describe mass and charge transfer in the ice/ice collision experiments of  \cite{Mason:00}.  This work is reviewed in \cite{Dash:06} and the basic mechanism has recently been adopted in a study of lightning in protoplanetary disks by \cite{Muranushi:2010}.   The \cite{Dash:01} theory focused on atmospheric ice rather than the very high pressures and very low temperatures that are encountered under nebular conditions, which substantially complicate the situation due to the detailed phase behavior of ice.  

Consider the surface between a solid ($s$) such as ice and the gas phase at a given temperature $T$ and pressure $P$. 
When one considers the field energy/molecule due to intermolecular attractions (polarization forces) across such a surface, there is a shift of bulk phase coexistence 
$\Delta\mu = \mu_{s}(T,P)- \mu_{\ell}(T,P)$ describing the formation of a stable interfacial  disordered or liquid film of thickness $d$ whose presence lowers the chemical potential of the wetted or disordered surface ${\mu}_{\cal I}(d)<0$.
 Hence,  this disordered equilibrium layer has a chemical potential $\mu_f$ that is {\em lower} than the {\em bulk} liquid $\mu_{\ell}$;   
$\mu_f (T,P,d) =  \mu_{\ell}(T,P) - |{\mu}_{\cal I}(d)| = \mu_{s}(T,P)$.  Furthermore, a variety of forms of disorder lower $\mu_f$; surface roughness, polycrystallinity, excess strain produced intrinsic or surface induced dislocations or Frank-Read sources, impurities or kinetic effects \citep{Dash:06}.  Such effects can be responsible for the persistence of disorder in at least a few molecular layers to ultra high vacuum temperatures. Importantly, all such forms of disorder can be present on the surface or within the bulk of nebular solids be they ice particles, growing from water vapor near the snow line, or warmer silicates in the inner nebula.  Although it is presently unfeasible to understand the detailed thermodynamic evolution of individual particles, the sign of the effect of disorder is the same for all.  Moreover, during interparticle collisions with speed $V_c$ the increased stress increases the density of disorder thereby further decreasing $\mu_f$ relative to the bulk liquid.  All of these effects are embodied as damage induced disorder, ${\mu}_{\cal D}(V_c)<0$, and  hence the combined influence of intermolecular forces, preexisting and damage induced disorder is to reduce the chemical potential of the interfacial liquid as  $\mu_f (T,P,d, V_c) =  \mu_{\ell}(T,P) - |{\mu}_{\cal I}(d)| - |{\mu}_{\cal D}(V_c)|$. 

Expanding $\Delta\mu$ about a point at coexistence, $T_m, P_m$ yields
\begin{equation}
 |{\mu}_{\cal I}(d)| + |{\mu}_{\cal D}(V_c)| = q_m \left(\frac{T_m - T}{T_m}\right) + \left(\frac{\rho_{\ell} - \rho_{s}}{\rho_{\ell} \rho_{s}}\right) (P_m - P), 
\label{eq:expansion}
\end{equation}
where $q_m$ is the latent heat of fusion, $\rho_{s}$ ($\rho_{\ell}$) is the density of the solid phase (liquid or high density phase), 
and for the usual form of nonretarded polarization forces $|{\mu}_{\cal I}(d)| = \frac{|{\cal A}_H|}{ \rho_{\ell} d^3}$, where ${\cal A}_H$ is the Hamaker constant divided by $6 \pi$.   ${\cal A}_H$ describes the strength of the polarization forces and is negative when the film is present \citep{Dash:06}.  Whence, a general thermodynamic relationship between the film thickness, the strength of the intermolecular forces and the damage induced energy density $u_{\cal D} \equiv \rho_{\ell} |{\mu}_{\cal D}(V_c)|$ is found as 
\begin{equation}
 d = \left[\frac{|{\cal A}_H|}{{\rho_{\ell}} \frac{q_m}{T_m}(T_m - T) + \left(\frac{\rho_{\ell} - \rho_{s}}{\rho_{s}}\right) (P_m - P) - u_{\cal D} }\right]^{1/3}. 
\label{eq:fullfilm}
\end{equation}
Consider for example that two particles of ice at fixed temperature $T$ collide.  The bulk sound speed in ice is $\sim 3 \times 10^5$ cm s$^{-1}$ and the thermal diffusivity is $\sim 5 \times 10^{-3}$ cm$^2$ s$^{-1}$ and thus only the interfacial region between the particles will relax thermally and mechanically and their remaining volume will dissipate the collision energy adiabatically. The rapid pressure build up during the collision ${P_c}$ drives the interfacial region towards equilibrium pressure at solid-liquid coexistence; $P_m(T)$.  Whence, both $T \rightarrow {T_m}^{-}$ and $P \rightarrow {P_m}^{-}$.  Simultaneously a fraction $\xi$ of the collisional energy  is converted into disorder enhanced interfacial damage $u_{\cal D}$ (some is lost through other channels such as e.g., surface waves and stored in other degrees of freedom) which lowers the chemical potential of the liquid phase.  Thus, the damage energy intervenes to further shift the onset of interfacial liquidity to values away from bulk coexistence.  Due to the rapid relaxation of the phonon modes in the interfacial region, thermodynamic state theory argues that, on the time scale of the collision,  the damage associated with the collisional energy density is deposited uniformly over the maximum collisional area $\pi {r_c}^2$ to a depth $d$ as $u_{\cal D} = \frac{\xi U_c}{\pi {r_c}^2 d}$, where $U_c = \frac{1}{2} {\cal M} {V_c}^2$ is the kinetic energy of the collision with ${\cal M} = {m_1 m_2}/({m_1 + m_2})$ the effective two-body mass. 

Analysis of Equation (\ref{eq:fullfilm}) for the range of astrophysically relevant particle sizes and collision energies shown in Figures \ref{fig:V(crit)} and \ref{fig:V(range)} reveals that it is  near the divergent limit, where the denominator vanishes, and hence 
\begin{equation}
 d = \frac{\xi U_c}{\pi {r_c}^2 \left[{\rho_{\ell}} \frac{q_m}{T_m}(T_m - T) + \left(\frac{\rho_{\ell} - \rho_{s}}{\rho_{s}}\right) (P_m - P) \right]}.  
\label{eq:film}
\end{equation}
Equation (\ref{eq:film}) shows that the detailed nature of the interactions responsible for the film become unimportant, but the form of the result depends on their existence in the first place.  
Thus, for example, for other power law interactions driving interfacial melting the exponent in Equation (\ref{eq:fullfilm}) changes and for exponentially decaying forces, such as are present due to screening in metals from conduction electrons or intrinsic or impurity produced ions in ice, the relation can be logarithmic  \citep{Dash:06}.  However, regardless of the nature of the intermolecular interactions, the film thickness diverges by the same competition of effects in the numerator of Equation (\ref{eq:fullfilm}) and hence Equation (\ref{eq:film}) is the same for all.  Therefore, while small changes in the nature of the attractive interactions between small particles, be they polarization forces or of some other form  \citep{Cowin:05}, may  influence the rate of growth of small agglomerations  \citep{Dominik:07, Blum:08, Guttler:10} below the \cite{Chokshi:93} limit   (see Figure \ref{fig:V(crit)}), they will have little effect here.

%\begin{figure}
%\begin{figure}[t]
%\centerline{\includegraphics[height=0.60\textwidth]{ProtoFigures/Fig1(a-b)_V1_Collide.pdf}}}}
%\hspace{0.2in}
%\centerline{\includegraphics[height=0.30\textwidth]{ProtoFigures/Fig1(a)_V1_Collide.pdf}}} 
%\centerline{\includegraphics[height=0.30\textwidth]{ProtoFigures/Fig1(b)_V1_Collide.pdf}}} 
%\hspace{0.65in}
%\centerline{\includegraphics[width=0.90\textwidth]{ProtoFigures/PTterms255-263.pdf}}
\begin{figure}
\centerline{\includegraphics[width=0.45\textwidth]{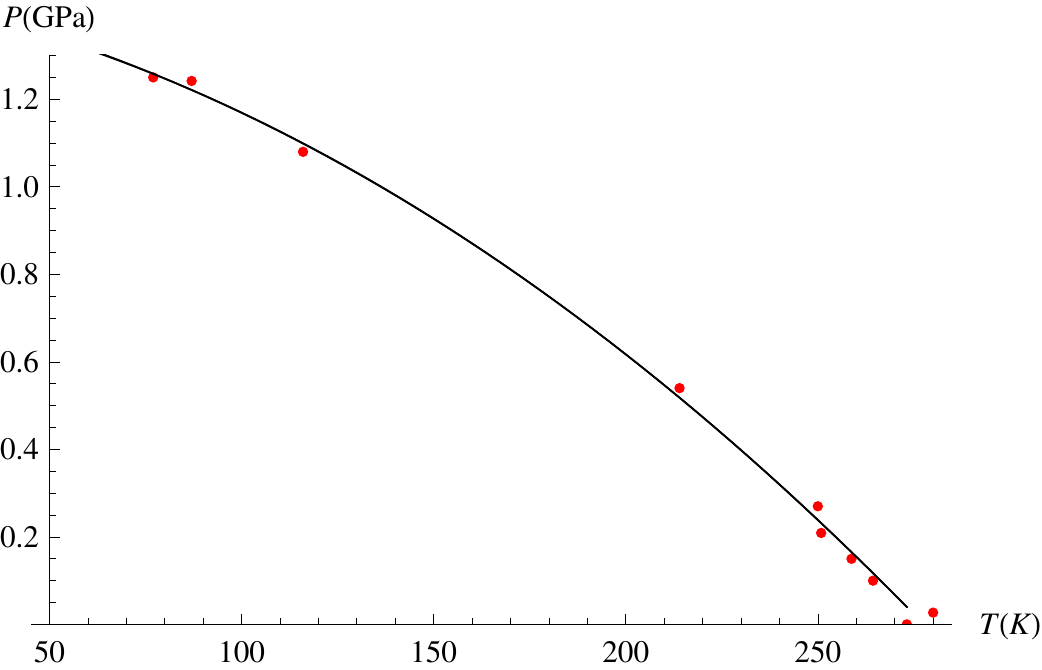}}
\caption{Phase Diagram: The equilibrium pressure $P_m=P_m(T)$ for ice particles at temperature $T$ is determined from a fit to the experimental data of  \cite{Straessle:07}.  The collisional energy density brings the interfacial region toward the high density disordered state as described by Equation (\ref{eq:film}). }
\label{fig:PhaseDiagram}
\end{figure}

\begin{figure}
%\begin{figure}[t]
%\centerline{\includegraphics[width=0.45\textwidth]{APJProtoFigures/StraesslePhaseDiagram.pdf}}
\centerline{\includegraphics[width=0.4\textwidth]{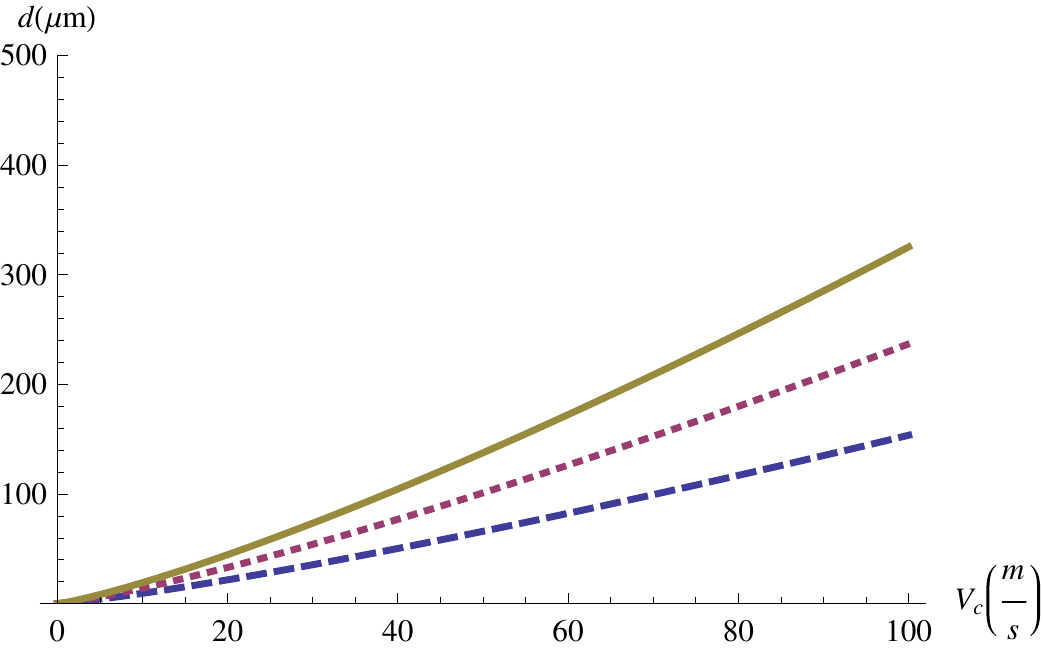}}
%\hspace{0.2in}
\centerline{\includegraphics[width=0.4\textwidth]{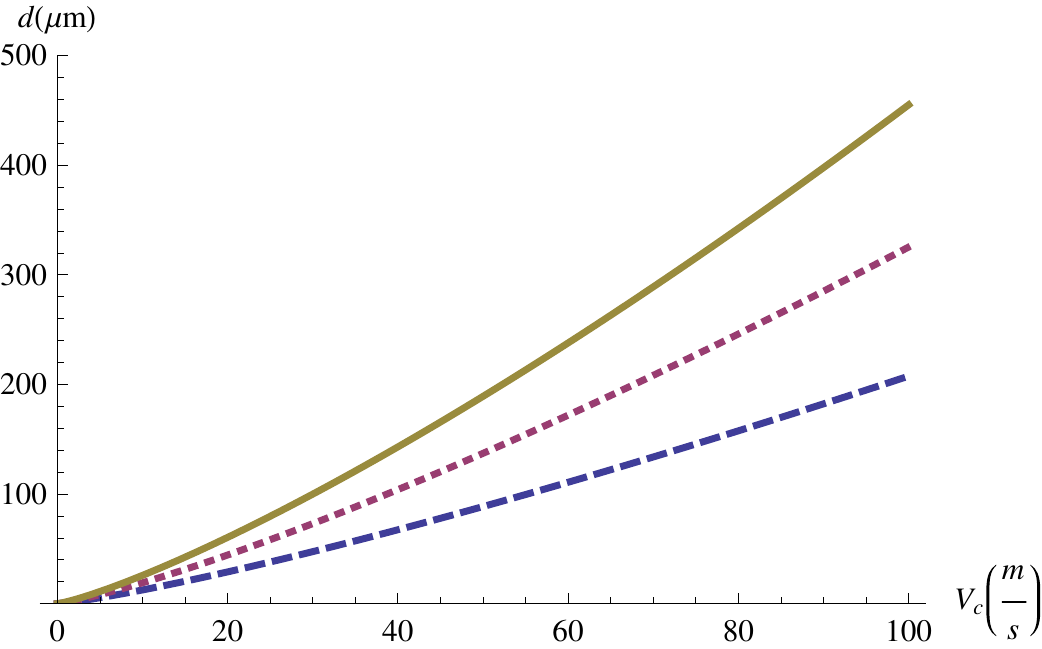}}
%\centerline{\includegraphics[height=0.30\textwidth]{APJProtoFigures/Fig1(a)_V1_Collide.pdf}}} 
%\centerline{\includegraphics[height=0.30\textwidth]{APJProtoFigures/Fig1(b)_V1_Collide.pdf}}} 
%\hspace{0.65in}
%\centerline{\includegraphics[width=0.90\textwidth]{APJProtoFigures/PTterms255-263.pdf}}
%\begin{figure}
\caption{Collisionally induced damage.  The  fraction $\xi$ of kinetic energy of the collision $U_c$ to interfacial liquid of thickness $d(\mu m)$ over collision area $\pi~{r_c}^2$.  In ice/ice central collisions between a 1 cm particle and a meter sized particle at a fixed temperature of (a) 105 $K$ and (b) 150 $K$. The solid ($\xi$ = 0.20), short dashed ($\xi$ = 0.15), and dashed ($\xi$ = 0.10) curves show the sensitivity to collision speed $V_c$ of the fractional conversion of energy to liquid.  As $T$ increases less collisional energy is required to create the same amount of liquid.  An increase in the energy density of the collision acts to bring the interfacial region toward the disordered (high density) liquid state.  \cite{Bridges:84, Bridges:96}, observed that rough frost covered ice spheres could readily stick during collisions.  Although their explanations and others following described the observation as some form of ``velcro'' at the surface, their findings are consistent with the idea here that an increase in the surface area results in a higher fraction of the energy being rendered on a more disordered region creating more melt.  The higher $T$ experiments of  \cite{Mason:00} show mass transfer in ice/ice collisions to be liquid like and their results depended on the degree of kinetic roughening which increases the surface roughness (``velcro'').  The abrupt increase in liquidity at the interface with $V_c$ is consistent with the findings of  \cite{Higa:98} of an abrupt decrease in the coefficient of restitution above a critical value of $V_c$ that increases as the radius of the small particle decreases.}
\label{fig:d(V)}
\end{figure}

\begin{figure}
\centerline{\includegraphics[height=0.30\textwidth]{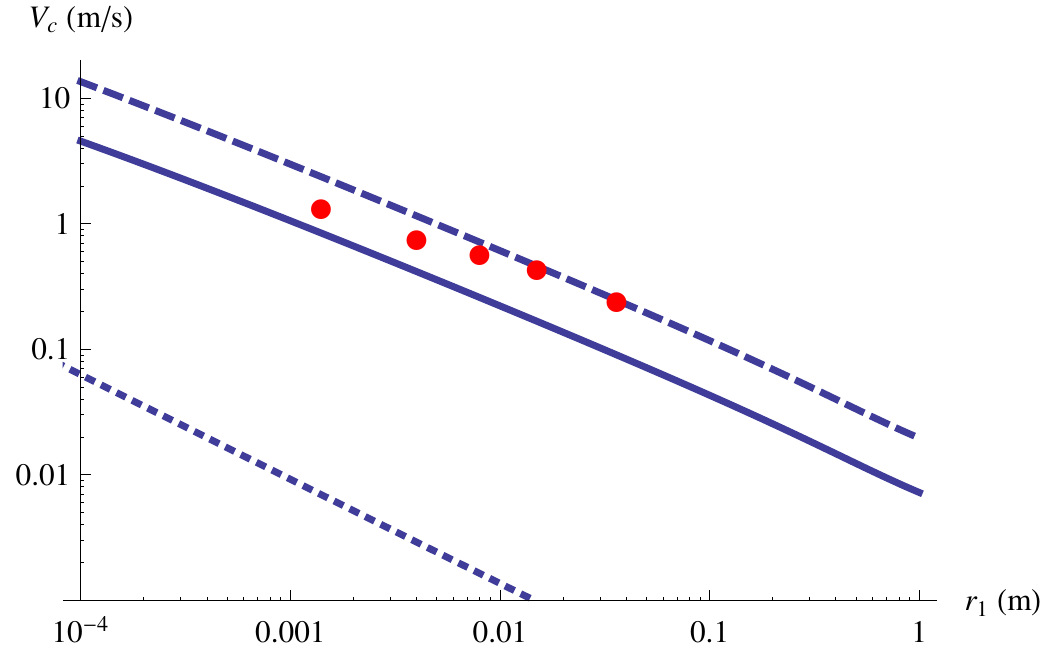}}
\hspace{0.2in}
%\caption{{\bf Collisional fusion.} 
\caption{ Constraining damage ($\xi$) in 
ice/ice central collisions, with speed $V_c$, between particles of radius $r_1$ and a particle of meter radius at a fixed temperature.    The solid curve is ${V_{L}}(r_1)$ ($\xi$ = 0.2) and the upper dashed curve is ${V_{U}}(r_1)$ ($\xi$ = 0.1).
The analysis represented in this figure is used {\em solely} to bound the range of damage ($\xi$ = 0.1-0.2) for sticking which is  used, in combination with the phase diagram of   \cite{Straessle:07} (Figure \ref{fig:PhaseDiagram}) at low temperatures and high pressures, to compute the astrophysical range of sticking beyond the snow line (Figure \ref{fig:V(range)}).
The points are from the experiments of  \cite{Higa:98} who measured an abrupt decrease to zero of the coefficient of restitution above a critical value of $V_c$ that {\em increases} as $r_1$ {\em decreases}.   While their experimental temperatures (261 $K$) are high relative to the typical snowline, they provide the widest available range of experimental parameters for relative particle radii and $V_c$.     
The ${V_{th}}$ threshold of \cite{Chokshi:93} is  the lower dotted curve, {\em below} which sticking is predicted because interfacial/intermolecular forces dominate elasticity. }
\label{fig:V(crit)}
\end{figure}

\begin{figure}
\centerline{\includegraphics[height=0.30\textwidth]{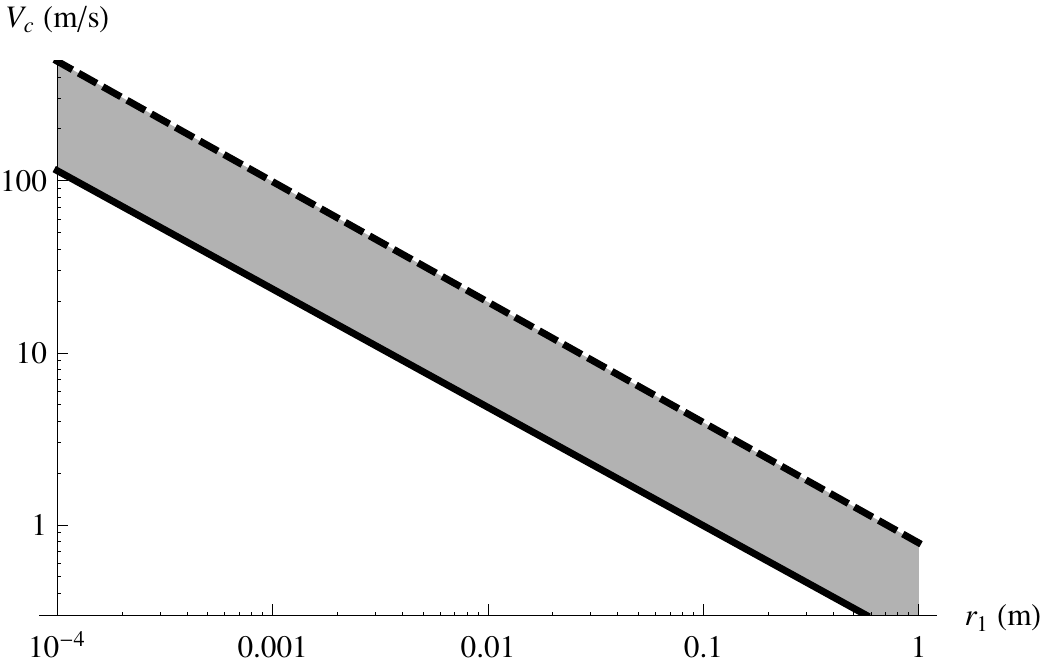}}
\hspace{0.2in}
\caption{
Calculated astrophysical collisional fusion range $\Delta V_c$ during ice/ice central collisions between particles of radius $r_1$ and a particle of meter radius at a fixed temperature using $\Delta \xi=0.1$ determined as described in Figure \ref{fig:V(crit)}.  The gray region between the lines delineates $\Delta V_c = {V_{U}}-{V_{L}}$ where sticking is operative.  The solid curve is ${V_{L}}(r_1)$ is for collisions at 150 $K$ (3.5 AU) and the upper dashed curve is ${V_{U}}(r_1)$ is for collisions at 105 $K$ (7 AU).  
The theory here predicts a new, high speed, regime of fusion that provides a physical basis for the high sticking coefficients necessary for rapid planetesimal growth overcome the bottleneck problem, as demonstrated in Figure \ref{fig:r(t)} below.  The range depends on the relative particle sizes and hence collisional energies, time scales,  the temperature, and thus position in the nebula.  We see that $\Delta V_c \sim$ 1-100 m s$^{-1} \gg {V_{th}}$ for ice.}
\label{fig:V(range)}
\vspace{0.2in}
\end{figure}

\subsection{Collisional Energetics: Fusion versus Bouncing}

The inelastic loss of energy into damage is treated as an inefficient Hertzian interaction.  Experiments show the Hertzian formalism to be accurate for the losses of up to 40\% of the incident kinetic energy  \citep{Gugan:2000}.  
Thus, 
the maximum value of the contact pressure $P\equiv \xi P_c$ and contact radius $r_c$ in Equation (\ref{eq:film}) are determined from
\begin{equation}
P_c = \frac{3}{2 \pi} \left(\frac{4}{3}\right)^{4/5} \left(\frac{5 {\cal M} {V_c}^2 E^4}{4 {\cal R}^3}\right)^{1/5}~~{\mbox{and}}
\end{equation}
\begin{equation}
r_c =  \left(\frac{15 {\cal M} {\cal R}^2 {V_c}^2}{16 E}\right)^{1/5} , 
\end{equation}
where ${\cal R} = {r_1 r_2}/({r_1 + r_2})$ is the effective two-particle radius and $E$ is the effective Young's modulus which, for particles of the same material and Young's modulus $\tilde{E}$ and Poisson's ratio $\nu$, is $E^{-1} = 2 (1 - \nu^2)/{\tilde{E}}$; for ice $E$ = 5.3 GPa.    
Figure \ref{fig:d(V)} shows that the interfacial liquid increases rapidly with $V_c$ and that this depends on the fraction  $\xi$ of the collisional energy channeled into disorder enhanced damage.

Particle  fusion requires a sufficient number of new bonds must be formed in the melting/freezing (amorph-polymorphization/annealing) process to overcome the stored elastic energy driving rebound and the liquid must freeze before separation. The combination of the degree of damage and these conditions determine ${V_{L}}$ and $ {V_{U}}$.  Collisions can clearly create sufficient interfacial liquidity (Figure \ref{fig:d(V)}) for fusion.  The
 interfacial liquid refreezes on a time scale ${\tau_{f}}$ commensurate with  the contact time $\tau_c$ and hence provides a velocity range for a given ${\cal R}$ in which fusion can occur.  The  contact time $\tau_c =  2.87\left({{\cal M}^2}/{{\cal R} V E^2 \xi^5}\right)^{1/5}$ is increased from the Hertzian value by the loss factor $\xi$ but is still dominated by the properties of the solid.  
  
In the limit that a thin annular region of liquid surrounding a sphere of solid is much thinner than the radius of the sphere, the time scale for freezing of the former by heat conduction through the cold solid is ${\rho_s q_m r_1 d}/{6 k_s \Delta T}$, 
where $k_s$ is the thermal conductivity of the solid and $\Delta T$ is the pressure corrected deviation of the melt fluid from the precollision bulk freezing temperature  \citep{Carslaw:1959}.   Under conditions for example at 150 $K$ the liquid formed is highly supercooled and thus likely freezes much more rapidly than this estimate which is very conservative.  More importantly, in laboratory experiments on freezing greater volumes ($\sim$ 30 mm$^3$ as opposed to the maximum volume here for the collisions at 150 $K$ of $\sim$ 20 mm$^3$ ) with $\Delta T$ of only approximately 30 $K$ is of order 10 ms \citep{Spannuth:07}.  Moreover, in the time scale estimate above the entire sphere is considered to be covered by the film so this estimate is an upper bound.  Nonetheless, for parsimony of development it is stated that the time it takes for the liquid formed by the damage to freeze is half that above due to heat loss from both sides of the interface; $\tau_{f} = {\rho_s q_m r_1 d}/{12 k_s \Delta T}$.   For cm sized particles colliding with a meter scale drifter  at $V_c \sim 10$ m s$^{-1}$ this gives  a range of $\tau_{f}$ for $T$ = 105 - 150 $K$ of 1-10 ms. Finally, as $T$ decreases  more energy is required for a given effective mass to create the same amount of liquid but $\tau_f$ decreases and hence the associated ${V_{U}}$ increases.  While these are rough estimates, a more refined analysis is not warranted by existing experimental data which do not cover the appropriate range of conditions.  The obvious refinements lead to decreases in $\tau_f$ and hence increase in both ${V_{L}}$ and ${V_{U}}$.  

There is a material specific interpretation of the contributions to ${\tau_{f}}$ and ${\tau_{c}}$ with a particular emphasis on the energetically favorable structure and hence the potential volume of fused material, $d\pi {r_c}^2$, which eventually drops below a value that provides sufficient bonding.   While  the collision time $\tau_c$ is interpreted as that of a weakly inelastic Hertzian process, in reality there is a continuous conversion of inelastic energy into the formation of interfacial melting throughout  $d\pi {r_c}^2$.  However (a) this conversion is extremely rapid since it is controlled by the phonon speed of the solid and (b) most of the elastic interactions are carried by the solid, except very near the divergent limit discussed below.  Thus, while being mindful of our degree of ignorance, it is understood that both bounds are likely  functions of the deviation of a collision from centrality, the surface history and angular momentum of the particles, among other unknown factors.  Finally, for ice there is an interesting but as yet unexplained size dependence of the maximum shear and tensile stresses   \citep{Higa:98} that may influence the slope of the upper curves in Figure \ref{fig:V(crit)}. 

\subsubsection{Collisions at the Divergent Limit}

Equation \ref{eq:film} shows that when the collisional energy is sufficiently high that the pressure and temperature reach their coexistence values $P_m$ and $T_m$ then $d$ diverges.   Firstly, it should be noted that over some ranges of collision speeds in liquid/liquid interactions of like material (e.g., water) collisions are still described quantitatively by Hertzian elasticity  \citep{Wang:08}, using the Laplace pressure as the ``Young's modulus'' viz., $\tilde{E}\approx\gamma/{\cal R}$ with $\gamma$ the liquid/vapor surface tension.  Whereas, when a liquid droplet collides with a superhydrophobic substrate the collision time is shown to be insensitive to $V_c$ which differs from Hertzian collisions  \citep{Richard:02} wherein the collisional energy is stored in the interfacial region.  Secondly, the liquidity here is indeed localized over an interfacial region of volume $ d \pi {r_c}^2$ that only diverges at a very large critical velocity; for a given $T$ and ${\cal R}$ there is an immeasurably narrow region of $V_c$ over which the Young's modulus of the small particle drops significantly below the large solid value.  Nonetheless, when including this effect here it is found only to be operative at values of $V_c \gg {V_{U}}$ and thus irrelevant.  Even were this not the case, the effect would be to {\em increase} $\tau_c$ and allow the thermal inertia of the larger particle more time to refreeze the smaller particle.  Thirdly, the heterogeneity and crystallinity of most materials is likely to lead to fracture at such high collisional energies.   Hence,  the rationale here is to simply increase the Hertzian value of $\tau_c$ by the loss factor $\xi$ due to the fact that in the collisional range of relevance the elastic interactions are principally borne by the solid.  

\subsection{\label{sec:Liquidity}Liquidity versus Structural Phase Transitions}

As described in section \ref{sec:Nebular} any calculation of the precise thermodynamic evolution of a given particle, and hence that of any particular two-body collision, is highly model dependent.
Here, the relevant phase behavior is described and the implications for other materials of particular relevance for the inner nebula are developed. 

At temperatures as low as 150 $K$ and high collisional pressures, ices Ih and Ic transform to supercooled water (likely rapidly through metastable ices II, III or IX)  \citep{Mishima:96, Straessle:07} whereas at lower temperatures LDA undergoes amorphization to HDA in the same modality as has recently been studied at high temperatures in H-passivated Si spheres   \citep{Suri:08}.  Rapidly deposited vapor may form as porous LDA but it will have ample time to anneal both thermally and due to cosmic irradiation  \citep{Palumbo:05} and, unless continuous rapid growth continues (which will allows the disorder to persist), such a particle will be either ice Ih or Ic upon collision.  Under the GPa pressures induced by typical high speed collisions these ices will, rapidly pass through the stable region of ice IX to transform into supercooled water  \citep{Mishima:96, Straessle:07}.  One can envision growth histories and outward drift trajectories to regions substantially lower than 150 $K$, \citep{Stevenson:88, Lecar:06, CieslaCuzzi:06, Garaud:07} and thus collisions between LDA ice, that can undergo an amorphization transition to HDA ice, or annealed crystalline ice, can collisionally melt by a phonon softening mechanism \citep{Straessle:07} represented by the phase diagram shown in Figure \ref{fig:PhaseDiagram}.  Regardless, the essential process discussed here is the same whether ice persists as LDA and then undergoes collisionally induced high pressure polyamorphization to HDA or anneals to Ih or Ic and collisions induce the formation of supercooled liquid  \citep{Mishima:96, Straessle:07}.  

The molecular dynamics simulations of \cite{Suri:08} provide an excellent example of damage induced polyamorphism and fusion in Si which experimentally melts at $\sim$ 1690 $K$.  They find that in high speed and hence high pressure collisions the $\beta$-tin phase forms in an interfacial region followed by picosecond annealing to the $a$-Si phase.  While their particles are sufficiently small that were they in the inner nebula they would be strongly coupled to the gas phase, 
the basic effect of low speed rebound ($V_c \sim$ 900 m s$^{-1}$) and high speed ($V_c \sim$ 1640 m s$^{-1}$) damage induced fusion is demonstrated. 
Clearly too, perhaps using simulations such as those on silicon \citep{Suri:08}, it is important to microscopically examine the role of the depth of disorder/liquidity necessary to fuse particles. 
While here material with laboratory determined properties is used, it is understood that agglomeration of micron scale particles can create highly porous ``pre-planetesimals'' in experiments and simulations \citep{Blum:08, Guttler:10}.  However, one must appreciate that over nebular time scales annealing can densify, and rapid growth kinetically roughen, all classes of crystalline material \citep{Dash:01, Dash:06}.  Note too that while porous materials may be more fragile, the higher roughness and porosity of non-annealed particles can be treated in this framework and the results may be qualitatively the same, but the preexisting disorder enhances the interfacial effects leading to fusion.  Thus, when wholesale fracture or spalling does not intervene, the mechanism described here provides a conservative range of collisional fusion.  

\section{Collisional Fusion and the Fate of Drifters}
\label{sec:Drift}

A particle  of radius $r(t)$ in the disk exchanges angular momentum with other solids and with the pressure supported sub-Keplerian nebular gas.  Whether the principal exchange of angular momentum is with the gas or the solids, one can derive equations of motion to predict the radial position of a particle in the disk $R(t)$ as a function of time.  When gas dominates then one needs a description of the gas density ${\rho_g}$ in the nebula, 
and when solids dominate their density $\rho_p$ determines the exchange process and hence the trajectory.  
While both density profiles are model dependent, a striking and robust result for the gas dominated exchange process, as first predicted by \cite{Adachi:76} and  \cite{Weidenschilling:77}, is that meter scale objects whose size does not change, rapidly spiral into the central star; for ${R(t=0)}$ = 1 AU it would take approximately 100 years for ${R(t)}$ = 0. 

In his study of  the dynamics of planetary rings \cite{Lissauer:84} showed how the addition of mass, and hence angular momentum, from meteoroids controls the specific angular momentum of the system. Thus,  as the material adjusts to conserve its total angular momentum there is a net inward drift.  \cite{Cuzzi:1993} studied these ideas in the context planetesimal growth using large scale numerical simulations in which they applied Reynolds decomposition to a two-fluid model (gas and solids) of a nebula.  They demonstrated that midplane conditions are such that the mass density of solids can substantially exceed that of the gas and yet turbulence still maintains conditions that do not allow the Safronov-Goldreich-Ward instability to drive gravitational collapse.  Therefore, while in general angular momentum exchange with both the gas and the solids  can drive a net inward drift, when $\rho_p \gg {\rho_g}$ solids dominate the angular momentum exchange and hence the drift rate.  Assuming perfect solid/solid sticking and assessing the angular momentum budget of drifters \cite{Cuzzi:1993} showed that their lifetime is dominated by the accretion of disk particles.   Here, the assumption of perfect sticking is abandoned and a simple analytical test of how the process of collisional fusion described here can influence the time evolution of accretion and hence the fate of a drifter is constructed as follows.  

When a drifting pre-planetesimal of radius $r(t)$ and mass $m$ receives additional mass by accretion of disk particles with average density $\rho_p$, it receives a net addition of angular momentum--a net torque--that is balanced by its inward motion. Writing the angular momentum  $\ell \equiv h m$ in terms of the specific angular momentum $h = {R} v_K \equiv \sqrt{ G M {R}}$ for a drifter in an orbit at ${R}$, where the central star has mass $M$, then 
\begin{equation}
\frac{\partial \ell}{\partial t} = {{\dot{m}}}(t) h + {\dot{R}(t)} m \frac{d h}{dR} = \rho_p \pi {r}^2 V_c (v_K - V_c){R}, 
\label{eq:ldot}
\end{equation}
and ${\dot{R}(t)}$ is the radial velocity through the disk.    The specific angular momentum $(v_K - V_c){R}$ contains a velocity $v_K - V_c$ measured relative to the (nearly Keplerian) gas.   Thus, the right hand side describes the net angular momentum evolution driven by the growth of the planetesimal from mass accretion at a rate $\rho_p \pi {r}^2 V_c$, which {\em assumes} perfect sticking/fusion for all collisional speeds $V_c$. Hence, it is seen that $4 \rho_s {{\dot{r}}}(t) = \rho_p V_c$, where $\rho_s$ is the internal density of the colliding particles, which, for example in the case of ice Ih is 917 kg m$^{-3}$.

Now,  the {\em bottleneck problem} is addressed here by considering such a midplane object with a trajectory dominated by mass accretion but modifying Equation \ref{eq:ldot} by (i) only accreting mass satisfying the collisional fusion criteria computed with the thermodynamic model; $\Delta V_c$ (Figure \ref{fig:V(range)}) and (ii) the midplane particle density $\rho_p(R,\alpha)$ will depend on position in the nebula $R$ and global turbulence characterized by $\alpha$ as described below.  

%\begin{figure}
%\begin{figure}[t]
%\centerline{\includegraphics[height=0.40\textwidth]{APJProtoFigures/Fig2(a-b)_V1_Collide.pdf}}}}
%\hspace{0.05in}
%\centerline{\includegraphics[height=0.30\textwidth]{APJProtoFigures/Fig2(a)235_04.pdf}}} 
%\centerline{\includegraphics[height=0.30\textwidth]{APJProtoFigures/Fig2(b)265_04.pdf}}} 
%\hspace{0.65in}
%\centerline{\includegraphics[width=0.35\textwidth]{PTterms255-263.pdf}}
%\centerline{\includegraphics[width=0.90\textwidth]{APJProtoFigures/PTterms255-263.pdf}}
\begin{figure}
\centerline{\includegraphics[width=0.4\textwidth]{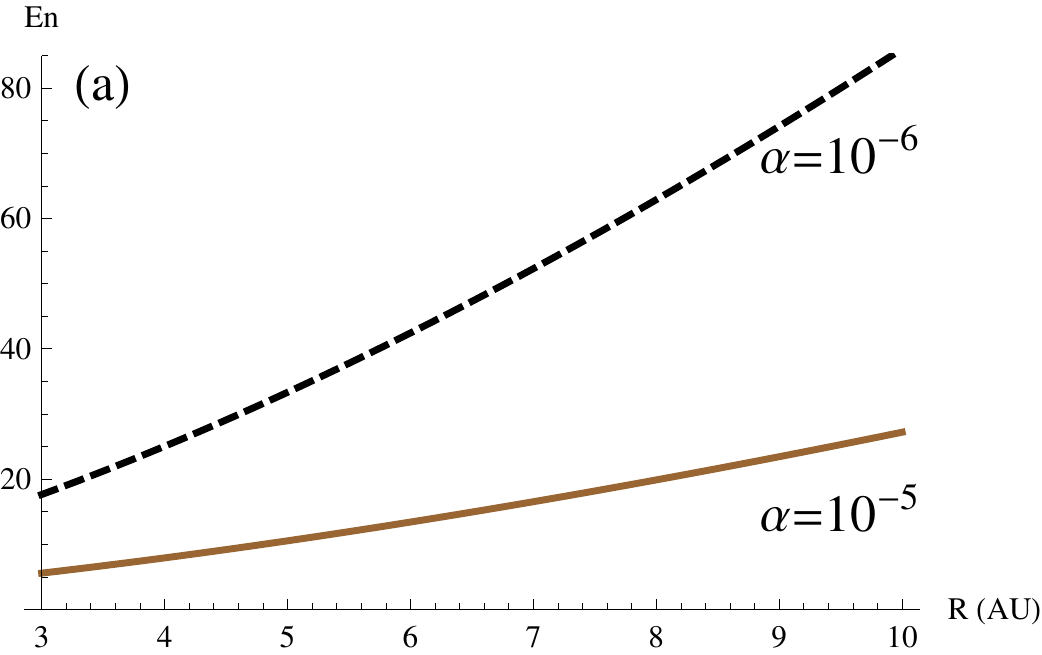}}
%\hspace{0.2in}
\centerline{\includegraphics[width=0.4\textwidth]{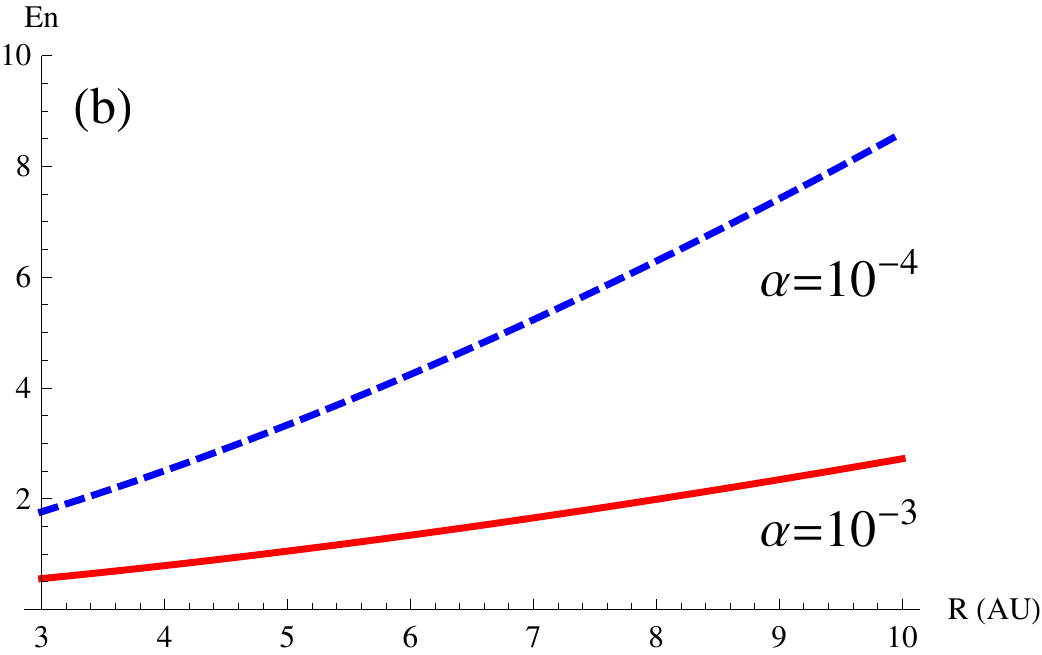}}
\caption{The midplane particle density enhancement factor  En $ \equiv {\rho_p}/{{\rho_g}}=10^{-2} {H}/{h_p}$ described by  Eqs (\ref{eq:En}) and (\ref{eq:scale}),  is plotted versus radial distance in the nebula $ R$ for 
(a) $\alpha = 10^{-6}$ and $\alpha = 10^{-5}$ (b) $\alpha = 10^{-4}$ and $\alpha = 10^{-3}$.  The mean free path in the gas is about 4 cm at 1 AU in this nebula, but exceeds a meter at about 3 AU well inside the region studied here for ice and hence the drifters are in the Epstein drag regime (where the particle radius is smaller than the mean free path) rather than the Stokes Regime (where the particle radius is larger than the mean free path).  
For a drifter with initial radius $r_{o}$ = 1 m, $St=St({R}, r_{o})$ ranges from 2 to 8 as $R$ ranges over the abscissa in the figure, but clearly as it grows by accretion to km scale, $St$ becomes exceedingly large.
Note that when treating collisions in the inner nebula, the number of the smallest particles undergoing central collisions will be reduced since aerodynamic effects will divert them on closest approach.  It is the enhancement of the midplane particle density,   $\rho_p(R,\alpha)= {\rho_g}({R})$ En,  that controls the collisional fusion driven growth rate of the drifter, $\dot{r}(t)$, thereby slowing its radial motion.  Clearly low turbulence disks exhibit an increase in $\rho_p$ and hence greater growth rates $\dot{r}(t)$.  
\bigskip}
\label{fig:En}
\end{figure}

A canonical cool ($T=280K$ at $R=1$AU) protoplanetary disk with an MMN gas to dust ratio of 100 \citep{Cuzzi:06, Armitage:09} is used and thus $\rho_p \sim 10^{-2} {\rho_g} H/h_p$, where ${\rho_g}(R) = 1.36 \times 10^{-9} R^{-11/4}$ g cm$^{-3}$ and I treat the influence of global turbulence on the gas ($H$) to particle ($h_p$) midplane layer vertical scale height ratio as follows.  Because the temperature $T$, sound speed $c$, and gas ${\rho_g}$  density all depend on radial position $R$, so too does the particle Stokes number $St(R,r) \equiv {t_s}{\Omega}$, where $t_s = t_s(R,r)$ is the stopping time of a drifter of or radius $r$ at position $R$ in the nebula where the rotational frequency is $\Omega(R)$. In the outer nebula, the gas mean free path ranges from  2 to 20 m for $3.4 \le R \le 10$ AU, and hence the Epstein drag regime is valid where $t_s(R,r) \equiv {\rho_s r(t)}/{{\rho_g (R)} c(R)}$. Thus, a drifter with initial radius $r_{o}$ = 1 m has 2 $\le St({R}, r_{o}) \le$ 8 for $3.4 \le R \le 10$ AU, but clearly as the accreting body grows to km scale, $St({R}, r)$ becomes exceedingly large. 
Because the rate at which the drifter moves through the disk 
${\dot{R}(t)}$ depends on the momentum drag via the fusional accretion rate of midplane solids, a simple representation of the $R$ and $\alpha$ dependence of $\rho_p$ is sought.  Global turbulence models \citep[e.g.,][]{Cuzzi:06} relate the gas and particle midplane layer vertical scale heights to the particle size dependent Stokes number $St({R}, r)$, and ${\alpha}$ as
\begin{equation}
\frac{H}{h_p} =  \sqrt{\frac{St(1+St)}{\alpha}}, 
\label{eq:scale}
\end{equation}
and thus a midplane particle density ``enhancement factor'' En=En$(R, St, \alpha)$ can be defined as 
\begin{equation}
\mbox{En} \equiv \frac{\rho_p}{{\rho_g}} = 10^{-2} \frac{H}{h_p}. 
\label{eq:En}
\end{equation}

\begin{figure}
\centerline{\includegraphics[width=0.4\textwidth]{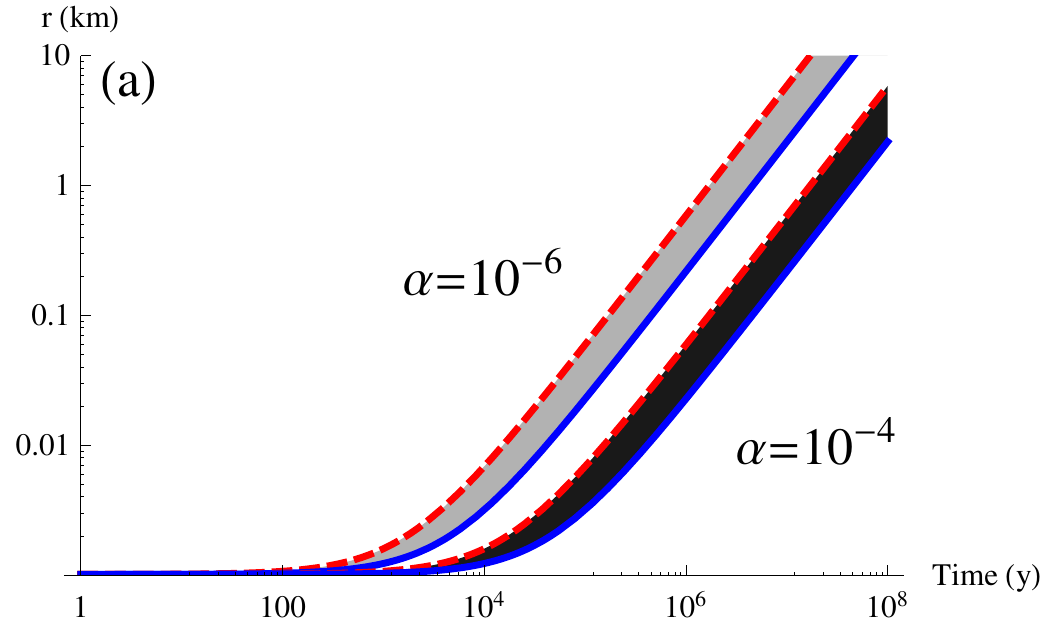}}
%\hspace{0.2in}
\centerline{\includegraphics[width=0.4\textwidth]{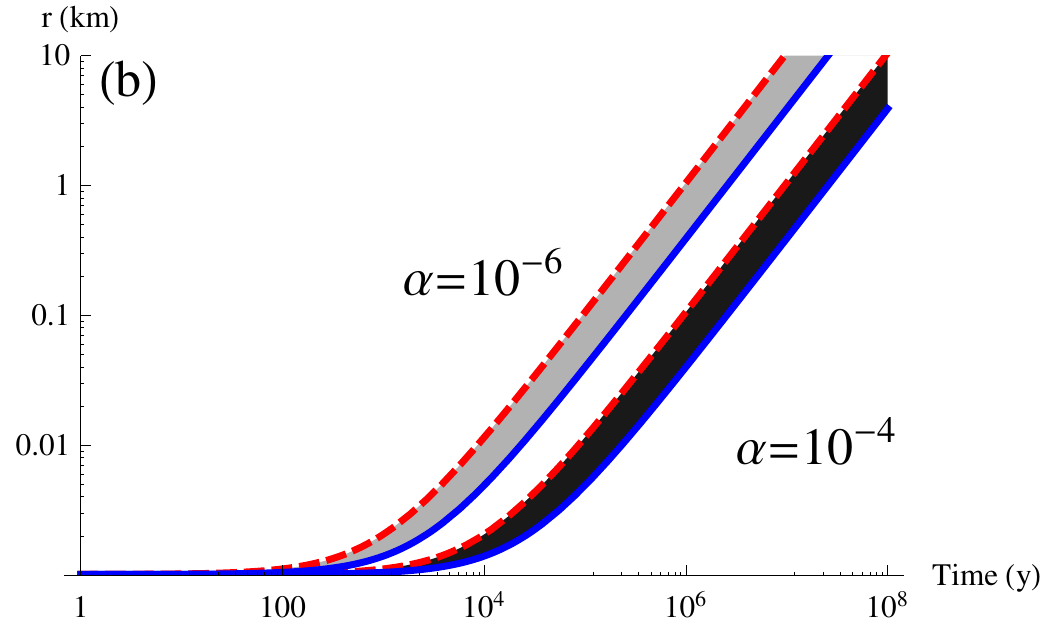}}
\caption{Drifting particle growth rates are
calculated using Equation (\ref{eq:r(t)}) at (a) $R_o$ = 7 AU where $T = 105 K$ and (b) $R_o$ = 3.5 AU where $T = 150 K$.   $\Delta V_c$ is determined from Figure \ref{fig:V(range)} giving the upper and lower bounds $V_U$ (red dashed) and $V_L$ (solid blue).
The  initial radius is a meter and results for $\alpha = 10^{-6}$ (light) and $10^{-4}$ (dark) are shown for a midplane particle size of a centimeter enhanced/suppressed  by  turbulence through En.  Figure \ref{fig:En} shows that increased turbulence decreases $\rho_p(R, \alpha)$ and hence there are fewer particles to fuse.  Whereas decreased turbulence increases ``momentum drag''  due to particle fusion and increases inspiraling, so that particles in a more laminar disk drift faster initially but also grow larger more rapidly and hence slow down more quickly as seen in Figure \ref{fig:DelR}.  The above figure can be compared to figure 19 of  \cite{Cuzzi:1993} which assumes perfect sticking.  Their smallest drifters were 10 m.  Here, even with the more restricted range of fusion $\Delta V_c$,  m scale drifters grow to km scale in less than $\sim$ Myr for weak turbulence and $\sim$ 0.1 Myr for much stronger turbulence. The essential point  is that drifters grow rapidly enough that they are not lost to the central star and thus remain to grow by gravitational attraction or other means (Figure  \ref{fig:DelR}). }
\label{fig:r(t)}
\end{figure}

\begin{figure}
\centerline{\includegraphics[height=0.25\textwidth]{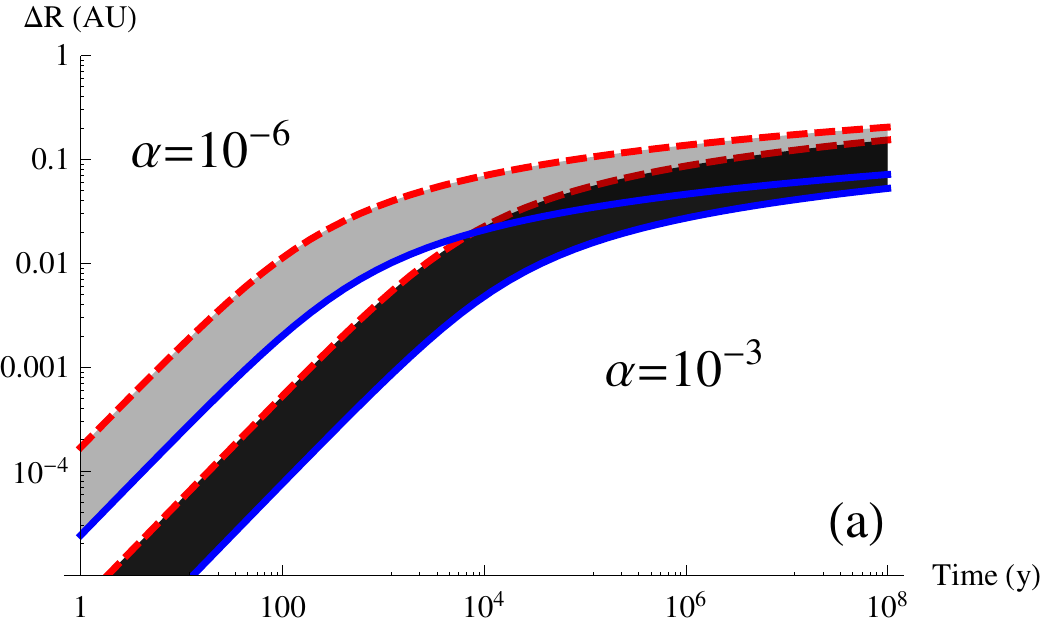}}
\centerline{\includegraphics[height=0.25\textwidth]{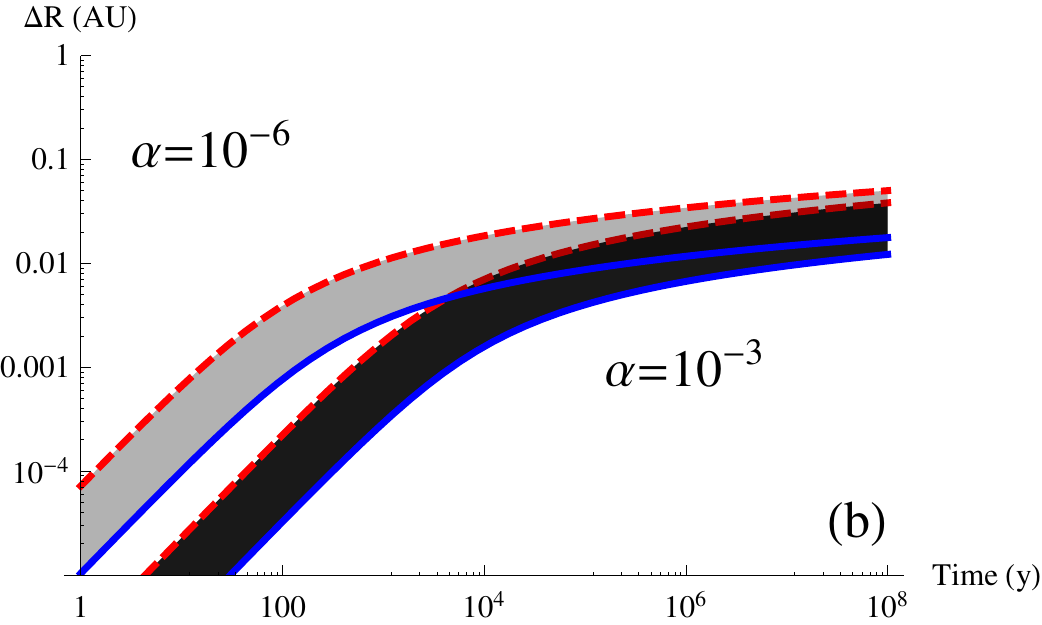}}
\caption{Decrease in orbital distance $\Delta R(t)$ due to collisional fusion and growth.   The collisions between any drifter and smaller particles in the midplane lead to accretion when the collision speed lies within $\Delta V_c$ according to the present theory.  By determining the general conditions for fusion Equation (\ref{eq:drift}) is solved using a turbulent enhancement/suppression of midplane density,  
(a) $R_o$ = 7 AU where $T = 105 K$ and (b) $R_o$ = 3.5 AU where $T = 150 K$;  well beyond and near
to the snow line.  Ice (or ice coated) drifters are treated across a wide range of temperatures and pressures.  
As described in the text, the range $\Delta V_c$  produces associated upper and lower bounds on the radial drift, ${\Delta R(t)}_U$ and ${\Delta R(t)}_L$, which are predicted for a variety of midplane particle sizes and turbulence enhanced/suppressed density $\rho_p(R, \alpha)$.   Results for $\alpha = 10^{-6}$ (light) and $10^{-3}$ (dark) are shown and do not change significantly regardless of whether the particles  in the midplane are a mm or 10 cm in size (shown here) or whether the drifter is a m or 100 m when it begins its journey.  Thus, it appears that this mechanism provides a firm basis for the previously assumed aspects of efficient sticking.}
\label{fig:DelR}
\end{figure}
 
The midplane enhancement factor is shown in Figure \ref{fig:En} beyond the snow line and clearly demonstrates that low turbulence disks have nearly a 100 fold increase in $\rho_p$.      The degree of enhancement of the midplane particle density  $\rho_p = \rho_p(R,\alpha)$  controls the collisional fusion driven growth rate of the drifter, $\dot{r}(t)$, and hence its radial motion $ {\dot{R}(t)}$.  Their trajectory in the disk depends on both their size (as they grow their radial speed decreases) and the particle drag (increases their inward drift in the same qualitative manner as does gas drag).  The growth rate $\dot{r}(t)$ an object of initial radius $r_{o}$ is 
\begin{equation}
\frac{{r}(t)}{r_{o} } 
= 1 + \frac{\rho_g(R) \mbox{En}V_c t}{4 \rho_s r_{o}},
\label{eq:r(t)}
\end{equation}
which provides the rate of mass accumulation on the right hand side of Equation (\ref{eq:ldot}) with $\rho_p=\rho_p(R, \alpha) = \rho_g(R) \mbox{En}$.  Figure \ref{fig:r(t)} demonstrates the effect of turbulent enhancement/suppression of $\rho_p$ on the accretion rate of drifters {\em controlled} by the thermodynamic constraints of collisional fusion.  Using an analytical estimate of the midplane particle density and removing the assumption of perfect sticking the ``momentum sweepup'' approach of \cite{Cuzzi:1993} is modified to provide a constrained calculation of ${\dot{R}(t)}$ as follows.
A drifter begins at $R(t=0) = R_o$ and {\em only} exhibits perfect sticking within the fusional bounds ${V_{U}} \ge V_c \ge {V_{L}}$ calculated from the general theory described above (Figure \ref{fig:V(range)}).   When particle drag dominates, the radial velocity through the disk ${\dot{R}(t)}$ of the growing pre-planetesimal is slowed by fusional accretion moving inward from $R_o$ to $R(t)$ and can be determined by integration of Equation \ref{eq:ldot} to be 
\begin{equation}
\Delta R(t) = \frac{6 V_c}{\Omega (R)} {\ell {n}}\left[\frac{{r}(t)}{r_{o} }\right], 
\label{eq:drift}
\end{equation}
where $\Delta R(t) \equiv R_o - R(t)$ and the argument of the logarithm is given by Equation (\ref{eq:r(t)}).   The consequences of the fusional bounds for drifters are examined in Figure \ref{fig:DelR} wherein 
one observes the upper and lower bounds on the radial drift; ${\Delta R(t)}_U$ and ${\Delta R(t)}_L$.   Figure \ref{fig:DelR} shows two examples of many that exhibit the same basic behavior using the new theory for collisional fusion to determine the sticking range of $V_c$, for a wide range of turbulent mixing of the disk midplane and thermodynamic conditions.  Figure \ref{fig:DelR}(a) shows ice drifters beginning at 7 AU at about 105 $K$, and \ref{fig:DelR}(b) at 3.5 AU and 150 $K$.  Particles evaporate at $\sim$ 170 $K$, when $R(t) \sim 2$ AU.  Interestingly, while decreasing turbulence increases the density of midplane target particles $\rho_p(R, \alpha)$, there is effectively higher drag and initially faster
inspiraling for lower turbulence because of angular momentum transfer.  Hence the particles in a more
laminar disk initially inspiral faster but also grow more rapidly.    One observes that at 
most, for $R_o$ = 7 AU, drifters move a few tenths of an AU before settling into stable Keplerian 
orbits and for $R_o$ = 3.5 AU it is an order of magnitude less.  While infrared observations of T Tauri disks are consistent with a nearly steady state distribution of particle sizes \citep{Dullemond:05}, the results in Figure \ref{fig:DelR} do not change in any significant way whether the particles in the midplane are a mm or 10 cm in size or whether the drifter is a meter or 100 meters when 
it begins its journey.  The principal point is that drifters are
not lost to the central star and so remain to grow by gravitational attraction and other means.

\section{Conclusion}
\label{sec:Conclude} 

Collisional fusion is a new mechanism for high velocity particle agglomeration.  It provides
a physical basis for the high sticking probability necessary, but heretofore assumed, to resolve the ``bottleneck'' problem in primary planetary accretion.   Thus, the radial drift of pre-planetesimals can indeed overcome the meter scale bottleneck based on a fundamental description of {\em how},  {\em why} and under what {\em dynamic} and {\em thermodynamic conditions} collisions with midplane particles can lead to perfect fusion.  
Operational regimes range from outer nebular regions, where collisions of ice, or ice covered, particles dominate and inner regions where higher melting temperature solids persist.  The frequency of central collisions in the latter will be reduced due to Stokes drag and would have to be incorporated in a detailed numerical model.  The process is qualitatively the same for cases in which the liquid phase is unavailable but a structurally favorable high pressure amorph/polymorph can be accessed during collisions; for water substance at low temperatures \citep{Mishima:96, Straessle:07}, or silicon at high temperatures \citep{Suri:08}.  In the case of liquifaction while some melt may be lost to the surroundings and is thus unavailable for fusion, under low vapor pressures it is known that splashing is suppressed \citep{Nagel:05}, whereas during polyamorphism this is obviously not an issue. While the range  $\Delta V_c \sim$ 1-100 m s$^{-1} \gg {V_{th}}$ calculated here captures astrophysical values \citep{Johansen:07}, we note that (i) the fracture strength of ice increases as the small particle radius decreases \citep{Higa:98}, so the relative size of the colliding particles is important, and (ii) even if perfect fusion does not occur the reduction in relative speed may be so dramatic that particles drop below the lower bound ${V_{th}} \sim$ 0.1-10 cm s$^{-1}$ of \cite{Chokshi:93}.  Hence, this new mechanism can act in concert with other mechanisms of particle destruction and sticking \citep{Guttler:10}.  For example, in the inner nebula, where Stokes drag is prevalent, the mechanism proposed here can coexist with that of \cite{Wurm:01} in which coupling to the gas phase is sufficiently strong that it can lead to reconnection of collisional fragments to a large object.  Therefore, the combined action of collisional fusion and shattering may underlie the perfect sticking required for rapid planetesimal accretion from the inner to the outer nebula.

\section{Acknowledgments} 

Support from Yale University, and the Wenner-Gren Foundation, the Royal Institute of Technology, and NORDITA in Stockholm is gratefully acknowledged.  The author benefited from discussions with A. Brandenburg, M.-M. Mac Low and E.A. Spiegel and the comments of the two referees. 

%\bibliographystyle{apj}
%\bibliography{Protocollisions}

\begin{table}
\centering
\setlength{\tabcolsep}{0.9 em}
\begin{tabular}{ll} \hline
\multicolumn{1}{l}{Symbol/Variable} & \multicolumn{1}{l}{Definition}  \\
\hline
$T, T_m$ & Temperature and Bulk Melting Temperature \\
$P, P_m$ & Pressure and Bulk Melting Pressure  \\
$\mu_q$ & Chemical Potential of quantity $q$\\
$q_m$ & latent heat of fusion or amorphization\\
$\rho_{s}$  & Mass density of the solid phase \\
$\rho_{\ell}$ & Mass density of the liquid or high density phase\\
%$\rho_q$ & Mass density of quantity $q$\\
%$\sigma_q$ & Surface density of quantity $q$\\
$V_c$ & Particle Collisonal Velocity  \\
$V_L$ ($V_U$) & Lower (Upper) Bound for Collisional Fusion\\
%$V_U$ & Upper Bound for Collisional Fusion\\
$\Delta{V_c}={V_{U}}-{V_{L}}$ & Range of Particle Collisonal Velocity for Fusion \\
$V_{th} \ll V_L$ & Lower Threshold of Particle Collisonal Velocity for Sticking \\
$U_c$ & Collisional Energy\\
$\xi$ & Fraction of Collisional Energy Converted to Damage\\
$u_{\cal D}$ & Damage Induced Energy Density\\
$d$ & Damage Induced interfacial film thickness\\
${\cal A}_H$ & Hamaker constant divided by $6 \pi$ \\
$E$ & Effective Young's Modulus\\
$\nu$ & Poisson's Ratio \\
$\cal M$ & Effective Mass of two-body collision\\
$\cal R$ & Effective Radius of two-body collision\\
$P_c$ &  Collisonal Pressure  \\
$r_c$ & Collisonal Contact Radius  \\
$\tau_c$ & Collision time\\
$\tau_f$ & Freezing/Annealing time for damage induced disorder\\
$k_s$ & Thermal conductivity of solid\\
$r(t)$ & Drifting particle radius as a function of time $t$\\
$r_{o}=r(t=0)$ & Drifting particle initial radius\\
$m(t)$ & Drifting particle mass as a function of time $t$\\
$h$ & Specific angular momentum of drifting particle\\
$\ell \equiv hm$ & Angular momentum of drifting particle\\
$R(t)$ & Nebular radial position as a function of time $t$.  $R(t=0) = R_o$.\\
$\Delta R(t) \equiv R_o - R(t)$ & Relative nebular radial position as a function of time $t$\\
$M$ & Mass of Central Star\\
$\Omega(R)$ & Orbital frequency $=2 \times 10^{- 7} R^{-3/2}$ with $R$ in AU\\
$G$ & Universal gravitational constant\\
$v_K\equiv\sqrt{GM/R}$ & Keplerian velocity\\
$\alpha$ & Global turbulence intensity parameter\\
${\sigma_p} ({\sigma_g})$ & Particle (gas) surface mass density in g cm$^{-2}$\\
${\rho_g}(R) =  1.36 \times 10^{-9} R^{-11/4}$ & Midplane MMN gas density in g cm$^{-3}$\\
${\rho_p}(R, \alpha) =  \mbox{En}~{\rho_g}(R)$ & Midplane particle density in g cm$^{-3}$\\
$H, h_p$ & Gas and particle vertical scale heights\\
$St=St(R,r_{o})$ & Particle Stokes number; spans the Epstein and Stokes regimes\\
En$(R)\equiv {\rho_p}/{{\rho_g}}=10^{-2} {H}/{h_p}$ & Midplane enhancement factor (Eq. \ref{eq:En})\\
$T(R) = T_0 (R/R_0)^{-1/2}$ & Vertically averaged radial nebular temperature; $T_0=280K$, $R_0$=1AU.\\
LDA & Low Density Amorphous Ice \\
HDA & High Density Amorphous Ice \\
\hline
\end{tabular}
\caption{List of Symbols and Variables}
\label{table:Variables}
\end{table}

\end{document}